\documentclass{emulateapj} 

\usepackage{longtable, hyperref, graphicx, subfigure, epstopdf, multirow, setspace, verbatim, lmodern} 

\usepackage[T1]{fontenc} 

\newcommand{\Msun}{\ifmmode {M_{\odot}}\else${M_{\odot}}$\fi} 
\newcommand{\Rsun}{\ifmmode {R_{\odot}}\else${R_{\odot}}$\fi} 
\newcommand{\lessim }{{\lower0.8ex\hbox{$\buildrel <\over\sim$}}} 
\newcommand{\gessim }{{\lower0.8ex\hbox{$\buildrel >\over\sim$}}} 
\newcommand\T{\rule{0pt}{2.6ex}}
\newcommand\B{\rule[-1.2ex]{0pt}{0pt}}
\def\amin{\ifmmode^{\prime}\else$^{\prime}$\fi} 
\def\asec{\ifmmode^{\prime\prime}\else$^{\prime\prime}$\fi}

\slugcomment{DRAFT \today} 

\shorttitle{Wide WDWDs from SDSS} 

\shortauthors{Andrews et al.} 

\begin{document} 

\title{COMMON PROPER MOTION WIDE WHITE DWARF BINARIES SELECTED FROM THE SLOAN DIGITAL SKY SURVEY}  

\author{Jeff J.\ Andrews\altaffilmark{1},  
Marcel A.\ Ag\"ueros\altaffilmark{1},  
Krzysztof Belczynski\altaffilmark{2,3},  
Saurav Dhital\altaffilmark{4,6},   
S.\ J.\ Kleinman\altaffilmark{5},  
Andrew A.\ West\altaffilmark{6}} 

\altaffiltext{1}{Columbia University, Department of Astronomy, 550 West 120th Street, New York, NY 10027, USA} 
\altaffiltext{2}{Astronomical Observatory, University of Warsaw, Al. Ujazdowskie 4, 00-478 Warsaw, Poland} 
\altaffiltext{3}{Center for Gravitational Wave Astronomy, University of Texas at Brownsville, Brownsville, TX 78520, USA} 
\altaffiltext{4}{Department of Physics and Astronomy, Vanderbilt University, 6301 Stevenson Center, Nashville, TN 37235, USA} 
\altaffiltext{5}{Gemini Observatory, Northern Operations Center, Hilo, HI 96720, USA} 
\altaffiltext{6}{Department of Astronomy, Boston University, 725 Commonwealth Ave, Boston, MA 02215, USA}

\begin{abstract} 

Wide binaries made up of two white dwarfs (WDs) receive far less attention than their tight counterparts. However, our tests using the binary population synthesis code {\tt StarTrack} indicate that, for any set of reasonable initial conditions, there exists a significant observable population of double white dwarfs (WDWDs) with orbital separations of 10$^2$ to 10$^5$ AU. We adapt the technique of Dhital et al.\ to search for candidate common proper motion WD companions separated by $<$10$\amin$ around the $>$12,000 spectroscopically confirmed hydrogen-atmosphere WDs recently identified in the Sloan Digital Sky Survey. Using two techniques to separate random alignments from high-confidence pairs, we find nine new high-probability wide WDWDs and confirm three previously identified candidate wide WDWDs. This brings the number of known wide WDWDs to 45; our new pairs are a significant addition to the sample, especially at small proper motions ($<$200 mas/yr) and large angular separations ($>$10\asec). Spectroscopic follow-up and an extension of this method to a larger, photometrically selected set of SDSS WDs may eventually produce a large enough dataset for WDWDs to realize their full potential as testbeds for theories of stellar evolution. 

\end{abstract}


\section{Introduction} 

Hydrogen-atmosphere (DA) white dwarfs (WDs), the evolutionary endpoints for stars with main-sequence masses between $0.8$ and $8\ \Msun$, are second in number only to low-mass main-sequence stars in the Solar neighborhood \citep{rowell11}. However, while they can remain at relatively high temperatures for Gyr, WDs are small and therefore usually faint objects, so that assembling complete catalogs of WDs has historically been challenging. The last decade has seen an impressive growth in our ability to find and characterize WDs, largely thanks to the Sloan Digital Sky Survey \citep[SDSS;][]{york00}. The first SDSS WD catalog included $2551$ WDs \citep{kleinman04}; the second nearly quadrupled that number, and included $6000$ new, spectroscopically confirmed WDs \citep{eisenstein06b}. The most recent SDSS WD catalog, based on the SDSS Data Release 7 \citep[DR7;][]{DR7paper}, contains $\sim$20,000 stars, of which 12,538 are classified as DAs (Kleinman et al., in prep.). 

Unsurprisingly, many of these WDs are in binary systems. The zoo of SDSS WD binaries is diverse; among its largest pens are those containing spectroscopic WD/M-dwarf pairs \citep[e.g.,][]{silvestri07,rebassa10,morgan12} and cataclysmic variables \citep[e.g.,][]{southworth10,paula11}. A smaller but especially interesting pen is the one for double WD (WDWD) systems. Depending on the composition and mass of the two WDs, WDWDs may be a formation channel for AM CVn-type binaries \citep[also found in SDSS;][]{anderson05,anderson08}, millisecond pulsars \citep[with the neutron stars formed through accretion-induced collapse; cf.\ discussion in][]{nelson05}, and Type Ia supernovae \citep[in the double-degenerate scenario;][]{webbink84,iben84,vankerkwijk10}. Furthermore, the discovery of WDWDs with mass ratios far from unity and with at least one WD with $M\ \lessim\ 0.2\ \Msun$ tests our understanding of common envelope evolution \citep{agueros09c,kilic10a,kilic11}.

Most known SDSS WDWDs are unresolved and/or compact binaries. By comparison, the expected population of wide, resolved WDWDs remains mostly unexplored, largely because of the observational challenges these types of binaries present, whatever their component stars. Wide binaries can have orbital periods that are much longer than human lifetimes. Members of these systems can only be identified astrometrically and confirmed through radial velocity measurements, although in practice the latter are hard to obtain for large samples of field stars.\footnote{Furthermore, typical line-of-sight velocities for WDs are comparable to the gravitational redshift experienced by the photons they emit \citep{silvestri01}.} Over his 70 year career, Luyten perfected a method that relies on proper motion ($\mu$) measurements to identify binary systems \citep[see for example the discussion in][]{luyten88}. The resulting pairs are known as common proper-motion binaries (CPMBs). This method is now commonly used and generally relies on high-proper-motion, magnitude-limited catalogs, so that searches for CPMBs tend to uncover nearby stars \citep[][and references therein]{dhital10}.

Wide binaries provide a unique perspective on stellar evolution: for those with orbital separations $a \gtrsim 10^2$ AU, the coeval components are far enough apart that mass exchanges are unlikely to have significantly impacted their individual evolutions \citep{silvestri01,fahiri06}. Wide binaries that include a main-sequence star and a WD can be used to determine the initial-to-final mass relation (IFMR) for WDs \citep[e.g.,][]{catalan2008a,zhao12}. The WD mass and cooling age are determined spectroscopically, and this age is subtracted from the main-sequence star's age to determine the WD's main-sequence lifetime and hence initial mass. WDWDs, for which total ages are not generally available, can still provide robust constraints on the IFMR \citep[e.g.,][]{finley97}. WDWD systems can also be used to constrain the effects of mass-loss on orbital evolution, for example by comparing their separations to the distributions for binaries containing two main-sequence stars and a main-sequence star and a WD \citep[e.g.,][]{sion91}. Unfortunately, the current sample of wide WDWD systems is small and heterogeneous, severely limiting its utility for these studies.

The surveys of \citet{giclas71} and \citet[e.g.,][]{luyten79} identified several thousand CPMBs; of these, a few hundred were thought to contain a WD. Spectroscopic follow-up allowed \citet{greenstein86} to confirm six as WDWDs and \citet{sion91} to increase that number to 21. Since then, individual WDWDs have been reported \citep[e.g., those found in the Palomar Green survey;][]{farihi05}, but there has been no systematic search for new wide WDWDs.

SDSS is an excellent dataset for identifying wide WDWDs. The photometric survey covered a large area to unprecedented depth \citep[$>$10$^4$ deg$^2$ and $\sim$$22$ mag;][]{DR7paper}, and matches between SDSS and USNO-B have been used to generate a proper motion catalog \citep{munn04} that is integrated into the SDSS database. Indeed, \citet{dhital10} used SDSS to uncover wide companions to low-mass main-sequence stars based on common proper motions: these authors identified over 1300 CPMBs, of which 21 include one WD.

We adapt the method developed by \citet{dhital10} to search for widely separated WD companions to the set of spectroscopically confirmed DA WDs included in the forthcoming Kleinman et al.\ catalog. In Section 2 we discuss the population synthesis predictions for the orbital distribution of WD binaries that motivated this search. We present in Section 3 our method for identifying candidate CPMBs and in Section 4 the properties of our newly discovered WDWDs; we also compare these properties to those of the previously known pairs. We conclude in Section 5.

\begin{deluxetable}{lr@{$\propto$}lc} 

\setlength{\tabcolsep}{0pt}

\tablewidth{0pt} 

\tablecaption{Initial Conditions for Population Synthesis  \label{tab:popsynth}} 
\tablehead{ 
\colhead{Parameter} &  
\multicolumn{2}{c}{Distribution} &
\colhead{Range} 
} 

\startdata 
Eccentricity & $\frac{dn}{de} \mbox{ }$ & $\mbox{ }2e$ & $0 < e < 1$ \\  
Primary IMF &  
$\frac{dn}{dM} \mbox{ }$ & $\mbox{} \left\{   
    \begin{array}{l} 
      M^{-2.2}\\ 
      M^{-2.7}\\ 
    \end{array}  
 \right.$  
& 
$ \begin{array}{l} 
\hspace{.1cm} 0.5 < \frac{M}{M_{\odot}} < 1 \\ 
\hspace{.1cm} 1 < \frac{M}{M_{\odot}} < 10 \\ 
    \end{array} $ \\ 
Mass ratio & $\frac{dn}{dq} \mbox{ }$ & $\mbox{ }1$ & $0 < q < 1$ \\ 
Orbital separation & $ \frac{dn}{da} \mbox{ }$ & $\mbox{ }a^{-1}$ & \hspace{.4cm} $ R_L < a < 10^5$ AU   
\enddata
\tablecomments{$R_L$ is the radius of the Roche lobe.} 
\end{deluxetable}

\section{Motivation} \label{motiv}

Whether a WDWD is observed today as a tight or a wide binary depends primarily on the orbital separation of the binary at birth. Because WD progenitors go through mass-losing giant phases, WDWD orbits may expand up to a factor of five due to conservation of angular momentum \citep{greenstein86}. If, however, the WD progenitors overflow their Roche lobes while in a giant phase, they will likely enter unstable mass transfer, causing a rapid, order-of-magnitude shrinking of the orbit \citep[e.g.,][]{iben93}. We therefore expect to observe a bimodal distribution of orbital separations, with a population of wide WDWDs that avoided any mass transfer phases and one of tight WDWDs that underwent unstable mass transfer. We use population synthesis to characterize this distribution in detail.

\subsection{Population Synthesis of WDWD Systems}\label{sec:pop_synth} 

Population synthesis is commonly used to analyze the formation and evolution of WDWDs. For example, \citet{nelemans01a} and \cite{nelemans05b} have used it to examine the physics of mass transfer, \citet{ruiter11} and \cite{meng11} to estimate SN Ia rates due to WDWD mergers, and \citet{ruiter10} and \citet{yu10} to predict the gravitational wave signal from inspiraling WDWDs  --- all phenomena related to tight WDWDs. Wide WDWDs, however, have largely been ignored in these studies.

Using the binary population synthesis code \texttt{StarTrack} \citep{bel02,bel08}, we evolve a sample of 10$^6$ zero age main sequence (ZAMS) binaries and determine the orbital separation distribution of the $\sim$5\% that become WDWDs. We summarize our initial conditions in Table~\ref{tab:popsynth} and describe them briefly below.

\begin{figure}[!th] 

\centerline{\includegraphics[width=.74\columnwidth,angle=90]{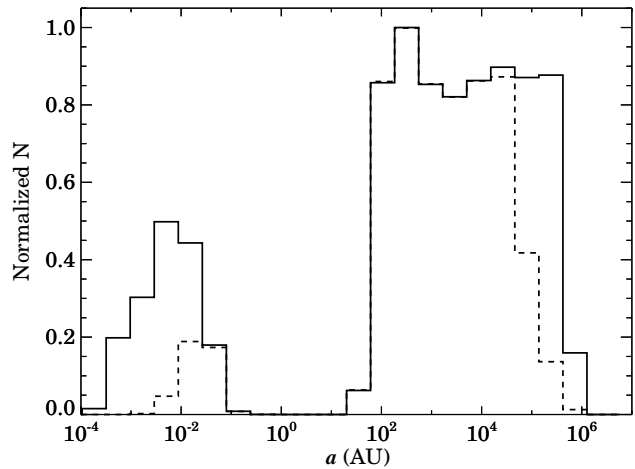}} 

\caption{Normalized orbital separation distribution of binaries produced by our population synthesis code once they become WDWDs (solid line) and after perturbing effects have been taken into account (dashed line). The compact systems are depleted due to mergers, while the widest systems are disrupted due to three-body interactions.}\label{Phist} 

\end{figure}

We use the standard Ambartsumian eccentricity distribution \citep{ambartsumian37,duquennoy91}. The initial mass of the more massive primary star is drawn from a Kroupa initial mass function \citep[IMF;][]{kroupa93}, and the initial mass of the companion star is a random fraction of the primary mass \citep{kobulnicky07}. Because of the weak dependence of the orbital separation on the component mass and orbital eccentricity, we expect that other (reasonable) distributions for these variables would not significantly alter the orbital separation distribution.

The initial orbital separations range from the two stars starting just outside of contact to being $10^5$ AU apart. The maximum separation of binaries at birth is not well constrained. However, $\sim$15\% of all G dwarfs are found in binary systems with separations $\gtrsim$$10^4$ AU \citep{duquennoy91}, the mean radii of pre-stellar cores are $\sim$$10^5$ AU \citep{clemens91}, and other studies have identified binaries with $a \sim 10^5$ AU \citep[e.g.,][]{dhital10}, so that this maximum initial separation is a reasonable estimate.

We choose a logarithmically flat distribution for $a$ \citep{opik24,poveda07}. Observationally, the distribution of birth orbital separations may be very different \citep[e.g.,][]{duquennoy91,cha04,lepine07}. However, when we test the broken power-law distribution of \citet{lepine07}, we see no significant differences in the resulting present day orbital separation distribution of WDWDs.

The solid line in Figure \ref{Phist} is the orbital separation distribution immediately after the birth of the second WD. As expected, this distribution is bimodal, with the number of systems with $a\ \gessim\ 10^2$ AU dominating the overall population. Interestingly, about 10\% of these wide pairs did experience mass transfer: the primary underwent stable Roche lobe overflow while on the asymptotic giant branch. In 90\% of cases, by contrast, the two stars can be considered to have evolved independently.

While the number and distribution of tight systems in Figure~\ref{Phist} is strongly dependent upon e.g., the prescription used to describe common envelope evolution, the existence of a population of a population of WDWDs widely separated at birth is a robust prediction of our code.

\subsection{Evolution of the Synthesized WDWD Orbits} 

WDWD orbits are modified over time by gravitational wave radiation and weak interactions with other bodies in the Galaxy. We use the equations of \citet{peters64} to model the first of these effects, which is most important for the tightest binaries in Figure \ref{Phist}. We model the second effect using the Fokker-Planck approximation, which determines the diffusion of energy into and out of the binary system. Using this approximation, \citet{weinberg87} found the characteristic lifetime $t$ of a binary consisting of two 0.6 $\Msun$ WDs to be: 

\begin{equation} \label{eq:lifetime} 
 t \approx 2.9 \left(\frac{a}{10^5\mathrm{AU}} \right)^{-1} \mathrm{Gyr}. 
\end{equation} 

The lifetime of a canonical WDWD with $a \sim 10^5$ AU is therefore a few Gyr.\footnote{This may be an overestimate of a wide binary's lifetime because it ignores the differential pull of the Galactic potential \citep{jiang10}. Although this effect is important, applying these authors' semi-analytic method to our population synthesis output is beyond the scope of this paper.}

We assign each ZAMS binary a birth time uniformly distributed over the lifetime of the Galactic disk (0-10 Gyr). The dashed line in Figure~\ref{Phist} shows the distribution of surviving binaries at the present day (10 Gyr), after gravitational radiation and Galactic interactions have caused some to merge and disrupted others. Even with our gross overestimate of the effect of Galactic perturbations, a large number of the wide WDWDs survive and should be observable today.

\begin{figure}[h] 

\centerline{\includegraphics[width=.73\columnwidth,angle=90]{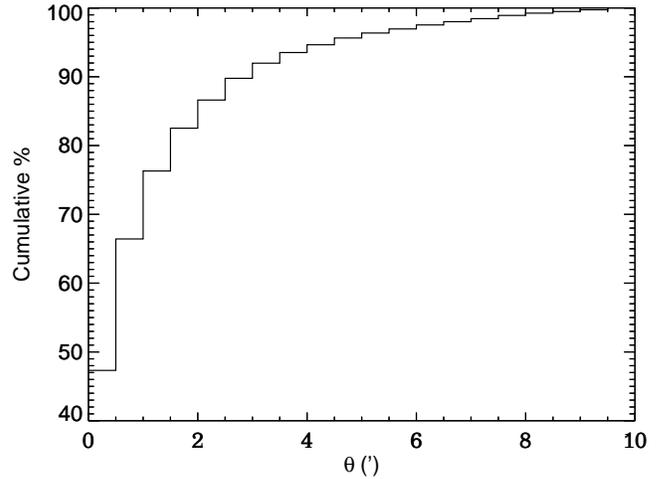}} 

\caption{Expected cumulative percentage of WDWD as a function of separation once projection effects are taken into account. The binaries are artificially placed at a distance of 250 pc. Close pairs dominate the distribution, but $\sim$15\% have $\theta\ \gessim\ 2$\amin.}\label{cdf_sep} 

\end{figure}

\subsection{Predicted Observed Angular Separation Distribution} 

To translate our synthesized population into an observable population, we assign each WDWD an argument of pericenter, mean anomaly, and cosine of the inclination angle, all randomly selected from  flat distributions. These parameters, combined with the intrinsic orbital parameters of the binary (eccentricity, masses, and orbital separation)  define the orbital positions of the two WDs. Using a fiducial distance of 250 pc, we then determine the projected angular separation ($\theta$) of each WDWD. Binaries with $\theta < 8$\asec\ or $>$10$\amin$ are eliminated because our method for identifying WDWDs (described below) is not sensitive to these. The predicted observed $\theta$ distribution of our synthesized WDWDs is shown in Figure~\ref{cdf_sep}. We find that the qualitative characteristics of this distribution are conserved for any reasonable assumption for the distance to the binaries. Although the WDWD population is dominated by pairs with separations $<$1\amin, a small fraction is expected to be observed at separations extending to 10$\amin$.

\begin{figure}[!t] 

\centerline{\includegraphics[width=0.75\columnwidth,angle=90]{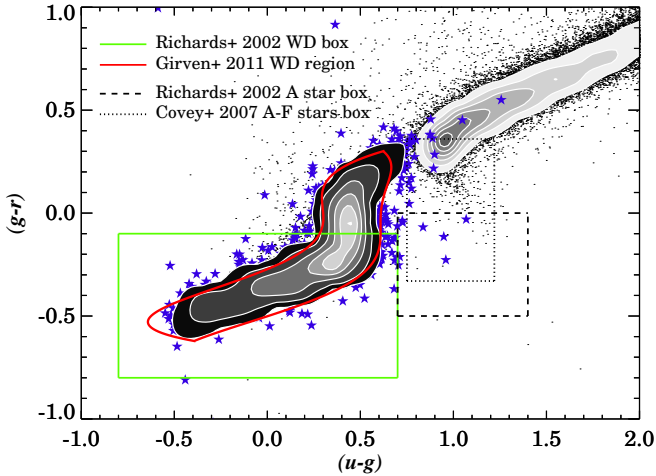}} 

\caption{$(g-r)$ vs.\ $(u-g)$ for 8605 spectroscopically confirmed DAs with a unique entry in the new Kleinman et al.\ catalog and with $ugriz$ photometric errors $\leq$0.15 mag (stars and dark contours). The points and light-colored contours to the upper right indicate the colors of $>$99,000 randomly selected stars with $ugriz$ errors $\leq$0.1 mag, $ugriz \geq$ 15.5 mag, and $g \leq$ 20 mag (these stars also meet the same proper motion constraints that were applied to our primary WDs). These stars are included to highlight the small overlap between the WD locus and the main sequence. Overplotted are different regions used to photometrically select WDs and A stars. The region defined by \citet{girven11} traces the empirical DA locus extremely well and includes 96\% of the 8605 DAs, a significantly larger fraction than returned by the standard \citet{richards02} color-cuts.}\label{cmd} 

\end{figure}

\section{Identifying Wide WDWDs in SDSS}  

\subsection{Candidate Binary Selection Process}\label{sec:winnow} 

The SDSS data reduction pipeline matches objects to objects in the USNO-B catalog, which has a limiting magnitude of $\sim$21 mag. Proper motions calculated during this matching are included in the \texttt{propermotion} table of the DR7 catalog.\footnote{Proper motions generated from this matching before DR7 contained a systematic error that has since been corrected \citep{munn08}.} We use the CasJobs database\footnote{\tt http://cas.sdss.org/casjobs/} to obtain all the available proper motion data for the DA WDs in the Kleinman et al.\ catalog: 11,563 DAs (92\%) have measured proper motions.

Following \citet{munn04}, we eliminate any WD with a rms fit residual $\leq$350 mas in either right ascension ($\alpha$) or declination ($\delta$) and any WD with more than one possible USNO-B counterpart (i.e., \texttt{match} $\ne$ 1). We also eliminate any WD with a total $\mu < 20$ mas/yr or with $\sigma_\mu > 10$ mas/yr in either coordinate. $\sim$56\% of the DAs in the Kleinman et al.\ catalog do not survive these cuts; most of these WDs have total proper motions below our threshold. We refer to the remaining $\sim$5500 DAs as primaries in our CPMBs and to their candidate companions as secondaries.

Next, we query CasJobs for all objects offset by $<$10$\amin$ from each of our primary WDs, yielding $\sim$$4.1 \times 10^6$ objects.\footnote{We require that these objects be defined by SDSS as primary, meaning they met a number of quality criteria that are described in \citet{stoughton02}.} Applying the proper motion quality cuts described above pares our list of potential secondaries down to $\sim$$2.2 \times 10^5$ objects. (Most of the eliminated objects lack proper motions only because they are too faint to be included in the USNO-B catalog.) We eliminate objects with poor photometry, i.e., with $g$-band errors $>$0.1 mag or $uriz$ errors $>$1.0 mag; this leaves $\sim$$2.0 \times 10^5$ objects.

We then search for objects with a proper motion matching that of our primary WDs. Typical proper motion observational errors are $\sim$4 mas/yr. In the most extreme case of a face-on orbit at the observational limit $\theta = 8\asec$ and at a distance of 50 pc, two WDs with typical $M = 0.6\ \Msun$ in a circular binary will have an orbital velocity of $\sim$0.8 km/s. This translates to a differential $\mu \sim 3.4$ mas/yr, which is of the same order as our observational $\sigma_\mu$. Since all our binaries are found at distances $>$50 pc (see below) and have $\theta > 8\asec$, the orbital velocities of our binaries can be ignored.

We define a proper motion match in the same manner as \citet{dhital10}: 

\begin{equation} \label{eq:match}
 \left(\frac{\Delta \mu_{\alpha}}{\sigma_{\Delta \mu_{\alpha}}} \right)^2 + \left( \frac{\Delta \mu_{\delta}}{\sigma_{\Delta \mu_{\delta}}} \right)^2 \leq 2, 
\end{equation} 

\noindent where $\Delta \mu$ is the scalar proper motion difference in $\alpha$ and $\delta$, and $\sigma_{\Delta\mu}$ is the error in the corresponding $\Delta \mu$, derived from the quadrature sum of the individual errors in $\mu$. This greatly reduces the number of candidate secondaries, to only 7129.

Finally, to identify candidate WDs among these secondaries, we select objects within 0.5$\sigma$ of the region in $(g-r)$ versus $(u-g)$ color-color space defined by \citet{girven11} as occupied by WDs. This region encompasses a larger area of the WD locus than do the standard \citet{richards02} color cuts. Furthermore, the \citet{girven11} region does not overlap with the $(g-r)$ versus $(u-g)$ boxes defined by \citet{richards02} and \citet{covey07} for main-sequence A and early F stars, the most likely stellar contaminants (see Figure~\ref{cmd}).\footnote{The separation between these regions is cleanest in this combination of colors.} 41 SDSS objects survive this final cut.

\begin{deluxetable}{lllll} 
\tablecaption{Candidate Wide WDWDs with SDSS Spectra for Both Components\label{tab:dist_rv}} 
\tablehead{ 
\colhead{} &  
\colhead{D$_1$} &  
\colhead{D$_2$} & 
\colhead{RV$_1$} & 
\colhead{RV$_2$} \\
\colhead{Name} &
\colhead{(pc)} &
\colhead{(pc)} &
\colhead{(km/s)} &
\colhead{(km/s)} 
} 
\startdata 
J0332$-$0049 & $143\pm21$ & $194\pm29$ & $-31\pm11$ & $-45\pm7$ \\ 
J0915$+$0947 & $216\pm32$ & $115\pm17$ & $-58\pm8$ & $+18\pm7$ \\ 
J1011$+$2450 & $436\pm65$ & $772\pm116$ & $+54\pm17$ & $+9\pm30$ \\ 
J1113$+$3238 & $147\pm22$ & $106\pm16$ & $+63\pm36$ & $+26\pm35$ \\ 
J1257$+$1925 & $516\pm77$ & $457\pm69$ & $+23\pm27$ & $+33\pm22$ \\ 
J1309$+$5503 & $137\pm21$ & $83\pm12$ & $-50\pm23$ & $+45\pm15$ \\ 
J1555$+$0239 & $291\pm44$ & $143\pm21$ & $+11\pm14$ & $-3\pm27$ \\ 
J2326$-$0023 & $115\pm17$ & $101\pm15$ & $-23\pm62$ & $-11\pm9$  
\enddata 
\tablecomments{These are all confirmed DA WDs included in either the \citet{eisenstein06b} or Kleinman et al.\ SDSS catalogs.} 
\end{deluxetable}

\subsection{Distances (and Radial Velocities)}\label{sec:distance} 

In addition to matching proper motions, bound binaries should have matching distances and (when available) radial velocities (RVs). For our candidate primaries, we use the spectroscopically derived $T_{\rm eff}$ and log $g$ values from the Kleinman et al.\ catalog and linearly interpolate $T_{\rm eff}$ and quadratically interpolate log $g$ in model evolutionary grids\footnote{{\tt http://www.astro.umontreal.ca/$\sim$bergeron/CoolingModels/}} \citep{tremblay11,holberg06,kowalski06} to determine their absolute magnitudes in each of the five SDSS bands, taking into account the best fit for Galactic reddening.\footnote{One candidate primary's $T_{\rm eff}$ and log $g$ values in the Kleinman et al.\ catalog were unusable because the code we use to generate these values rejected the associated calibrations. We obtain $T_{\rm eff}$ and log $g$ estimates for this WD by redoing the fit (K.\ Oliveira, private communication).} The distance to each primary is then the average of the distances derived in each band.\footnote{Fits to the grids also provide cooling ages and masses.} Formally, the uncertainties associated with these distances are very small (a few percent). However, these do not reflect the hard-to-quantify uncertainties in the model grids, and we therefore adopt a more realistic uncertainty of 15\% for these distances.

Determining photometric distances to our candidate secondaries, which generally lack SDSS spectroscopy, is less straightforward. While $T_{\rm eff}$ can be derived from SDSS photometry to within a few 100 K, the typical photometric errors are sufficient to create uncertainties of order half a decade in the derived log $g$. As a result, one typically assumes log $g = 8.0$ (the Kleinman et al.\ DA sample is strongly peaked at this value). We compare the distances calculated in this manner for our candidate primaries to those derived from their spectroscopically determined $T_{\rm eff}$ and log $g$ values. We find that they are consistent only to within a factor of 2. In general, therefore, the photometric distance uncertainties are too large for us to differentiate between real and spurious binaries based on comparisons of the distance to the two stars.

There are eight candidate pairs in which both components have SDSS spectra. We provide the spectroscopically derived distances and cataloged RV measurements for these WDs in Table~\ref{tab:dist_rv} and discuss these systems in more detail in Section \ref{res}.

\subsection{Estimating the Purity of Our Sample} 

We use two methods to test the robustness of our selection. First, we estimate the overall contamination of our sample by shifting the positions of our primary WDs and using the same photometric and proper motion criteria to identify (false) companions to these shifted primaries. Second, we use the Monte Carlo approach developed by \citet{dhital10} to estimate how likely it is to find a random star whose characteristics match those of our candidate primaries.

\begin{figure} 

\centerline{\includegraphics[width=0.74\columnwidth,angle=90]{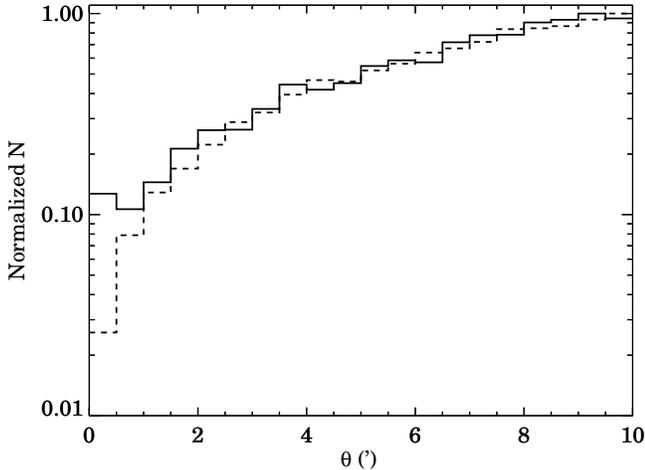}} 

\caption{$\theta$ distribution of real (solid line) and of false (dashed) candidate CPMBs. The primary WD positions were shifted by $\pm$1$^\circ$ in both $\alpha$ and $\delta$; we then applied our method for identifying common proper motion companions to these shifted stars. There is an excess in the distribution of real candidates for $\theta \leq 1.5-2$\amin; at larger $\theta$, the contamination by false pairs is essentially 100\%.}\label{hist_sep} 

\end{figure}

\subsubsection{Empirical False Positive Determination}\label{sec:false_pos} 

We shift the positions of our primary DA WDs four times ($\pm$1$^\circ$ in both $\alpha$ and $\delta$) and use the $\mu$ cuts described above to identify candidate proper-motion companions to these stars. In Figure~\ref{hist_sep}, we compare the normalized $\theta$ distribution of the resulting population of false candidate CPMBs after these shifts and that of the 7129 candidate CPMBs identified when using the true DA positions. There is an excess in the distribution of real candidates for $\theta \leq 1.5-2\amin$, while at larger separations, the distributions are equivalent, suggesting that most of our actual candidates are due to random matches.

\begin{figure} 

\centerline{\includegraphics[width=0.74\columnwidth,angle=90]{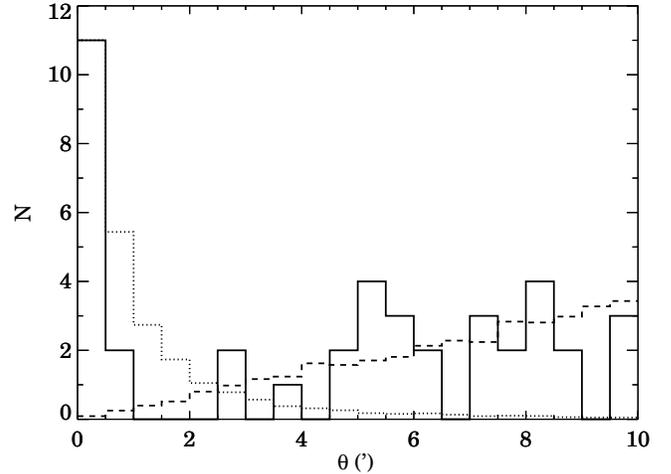}} 
\caption{$\theta$ distribution of the $41$ actual WDWD candidates (solid line) compared to that of the artificially generated candidate CPMBs (dashed line). The dotted line is the predicted distribution from population synthesis (shown in Figure~\ref{cdf_sep}) normalized to the first bin in the real distribution, and suggests that our sample is incomplete at $\theta < 2\amin$. Our distribution of real candidates includes more candidates at large $\theta$ than is expected from our population synthesis results, and is strongly contaminated for $\theta > 2\amin$.}\label{hist_wd} 

\end{figure}

In Figure~\ref{hist_wd}, we compare the $\theta$ distribution for the candidate WDWDs selected from the real CPMB candidates to that for the false candidate CPMBs shown in Figure~\ref{hist_sep}. Given the small number of candidate WDWDs, we also plot the predicted distribution for WDWDs from Figure~\ref{cdf_sep} normalized to the first $\theta$ bin in our distribution of real candidates. While Figure~\ref{hist_sep} suggests that true binaries are most likely at separations $\leq$$1\amin$, the total expected number of observed wide WDWDs (dotted line in Figure~\ref{hist_wd}) is significantly larger than the expected number of false positives (dashed line) out to $\theta \sim 2\amin$. Although none of our actual candidates have $1\amin< \theta < 2 $\amin, our results suggest that such WDWDs, if found in future searches using similar constraints, are likely to be true binaries.

\subsubsection{Galactic Model}\label{sec:gal_mod} 

Each realization of the \citet{dhital10} Galactic model populates a 30$\amin$ $\times$ 30$\amin$ conical volume centered at the position of the primary WD up to distances of 2500 pc from the Sun. The model assigns a position in six-dimensional phase space to each simulated star, assuming three kinematic components of the Galaxy corresponding to thin disk, thick disk, and halo populations.

After $10^5$ realizations of the model, we count the number of rendered stars for which $\mu$ matches $\mu_{primary}$ (as defined by Equation~\ref{eq:match}) and that are at a separation from the primary smaller than or equal to the separation of the corresponding candidate binary. A matching star also has to be at a distance consistent with the distance to the primary WD to within the quadratic sum of the distance uncertainties to the primary and secondary in the binary being tested. In most cases, this corresponds to searching a relatively large volume along the line-of-sight for matches, since while the spectroscopic distances uncertainties are 15\%, the photometric distance uncertainties are taken to be 100\%. For the eight primaries whose secondaries also have spectroscopic distances and RV measurements (see Table~\ref{tab:dist_rv}), the searched volume is smaller, and we further require that a match have an RV consistent with that of the primary WD to within $1\sigma$. Unsurprisingly, this results in a systematically lower number of random matches to the primaries in these pairs.

\begin{figure} 

\centerline{\includegraphics[width=0.75\columnwidth,angle=90]{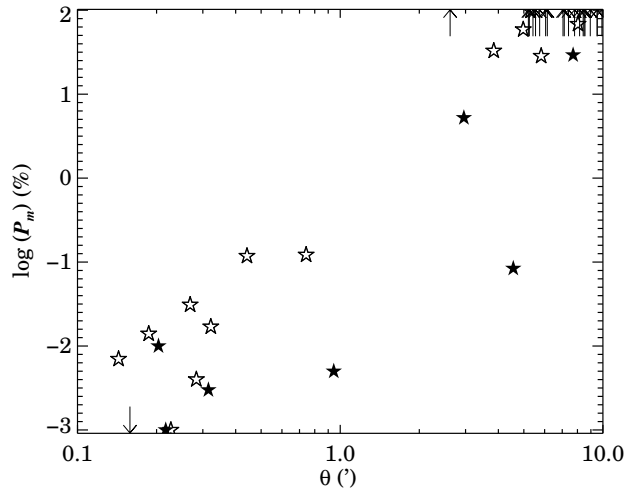}} 
\caption{Percentage of model realizations that return a random star with the same properties as those of the $41$ primary WDs in our candidate WDWD binaries; $\theta$ is the separation of the corresponding candidate WDWDs. The arrows indicate WDs for which $P_m$ is either less than or greater than the y-axis limits. $P_m$ can be $>$100\% because a given realization may have more than one star with properties matching those of the candidate being tested. Note that there is a primary at $\theta \sim 5\amin$ with $P_m \sim 100\%$ and two primaries at $\theta \sim 8\amin$ that are also partially covered by the arrows. 13 primaries have $P_m < 1\%$ and $\theta < 2\amin$.}\label{gal_mod} 

\end{figure}

Figure~\ref{gal_mod} shows the percentage $P_m$ of realizations that return a random star whose properties match those of our $41$ candidate primary WDs. These results are in agreement with those from our first test: finding a random match within 2\amin\ of one of our primary WDs is extremely unlikely, while pairs with $\theta\ \gessim\ 5\amin$ are much more likely to be random matches. Interestingly, however, this test suggests that there are systems with $2\amin < \theta < 5\amin$ that are more likely to be real than random, as the likelihood of finding a random star with properties matching those of the WD primaries is small.

\begin{deluxetable*}{llllllllll} 
\setlength{\tabcolsep}{0.005in}  
\tablecaption{High-Confidence Candidate Wide WDWDs \label{tab:results}} 
\tablehead{ 
\colhead{Name} &  
\colhead{$\alpha$} &  
\colhead{$\delta$} & 
\colhead{$g$ (mag)} &  
\colhead{$\mu_{\alpha}$ (mas/yr)} &  
\colhead{$\mu_{\delta}$ (mas/yr)} & 
\colhead{Type} & 
\colhead{$\theta$ ($\asec$)} & 
\colhead{$P_m$} & 
\colhead{Ref.} 
} 
\startdata 
\multirow{2}{*}{J0000$-$1051} & 00:00:22.5 & $-$10:51:42.1 & $18.91\pm0.04$ & $+45.3\pm4.6$ & $-25.3\pm4.6$ & DA & \multirow{2}{*}{16.1} & \multirow{2}{*}{$<$0.1\%} & 1,2 \\
 & 00:00:22.8 & $-$10:51:26.6 & $20.21\pm0.04$ & $+42.3\pm4.1$ & $-24.3\pm4.1$ & & & & \B \\
\hline
\T
\multirow{2}{*}{J0029$+$0015\tablenotemark{a}} & 00:29:25.6 & +00:15:52.7 & $18.48\pm0.03$ & $-28.9\pm3.1$ & $-23.0\pm3.1$ & DA & \multirow{2}{*}{8.6} & \multirow{2}{*}{$<$0.1\%} & 1,2 \\
 & 00:29:25.3 & +00:15:59.8 & $19.59\pm0.02$ & $-27.9\pm3.6$ & $-23.9\pm3.6$ & & & & \B \\  
\hline
\T
\multirow{2}{*}{J0332$-$0049} & 03:32:36.6 & $-$00:49:18.4 & $18.20\pm0.03$ & $-24.8\pm5.4$ & $-23.6\pm5.4$ & DA & \multirow{2}{*}{18.9} & \multirow{2}{*}{$<$0.1\%} & 1,2,3,4 \\
 & 03:32:36.9 & $-$00:49:36.9 & $15.64\pm0.02$ & $-30.9\pm4.5$ & $-23.3\pm4.5$ & DA & & & 2 \B \\
\hline
\T
\multirow{2}{*}{J1002$+$3606} & 10:02:44.9 & +36:06:29.5 & $18.92\pm0.03$ & $-32.9\pm3.3$ & $-27.5\pm3.3$ & DA & \multirow{2}{*}{26.5} & \multirow{2}{*}{0.12\%} & 1,2 \\
 & 10:02:45.8 & +36:06:53.3 & $19.04\pm0.02$ & $-29.9\pm3.4$ & $-27.0\pm3.4$ & & & & \B \\
\hline
\T
\multirow{2}{*}{J1054$+$5307\tablenotemark{b}} & 10:54:49.9 & +53:07:59.2 & $17.92\pm0.04$ & $-113.9\pm3.1$ & $-38.4\pm3.1$ & DA & \multirow{2}{*}{44.5} & \multirow{2}{*}{0.12\%} & 1,2,4,5 \\
 & 10:54:49.2 & +53:07:15.2 & $17.52\pm0.03$ & $-112.9\pm2.9$ & $-36.0\pm2.9$ & DA & & & 4,5 \B \\
\hline
\T
\multirow{2}{*}{J1113$+$3238\tablenotemark{c}} & 11:13:19.4 & +32:38:17.9 & $19.03\pm0.03$ & $-162.9\pm3.1$ & $+58.0\pm3.1$ & DA & \multirow{2}{*}{56.7} & \multirow{2}{*}{$<$0.1\%} & 2 \\
 & 11:13:22.6 & +32:38:58.9 & $19.12\pm0.04$ & $-158.8\pm3.2$ & $+58.0\pm3.2$ & DA & & & 2 \B \\ 
\hline
\T
\multirow{2}{*}{J1203$+$4948} & 12:03:11.5 & +49:48:32.4 & $19.03\pm0.03$ & $-97.6\pm3.4$ & $-36.5\pm3.4$ & DA & \multirow{2}{*}{19.3} & \multirow{2}{*}{$<$0.1\%} & 2 \\
 & 12:03:11.0 & +49:48:50.8 & $17.35\pm0.02$ & $-98.2\pm2.9$ & $-39.1\pm2.9$ & & & & \B \\
\hline
\T
\multirow{2}{*}{J1257$+$1925} & 12:57:20.9 & +19:25:03.7 & $19.88\pm0.06$ & $-38.4\pm5.4$ & $-31.6\pm5.4$ & DA & \multirow{2}{*}{12.2} & \multirow{2}{*}{$<$0.1\%} & 2 \\
 & 12:57:21.1 & +19:24:51.8 & $17.07\pm0.03$ & $-33.0\pm2.7$ & $-33.5\pm2.7$ & DA & & & 2 \B \\
\hline
\T
\multirow{2}{*}{J1412$+$4216\tablenotemark{d}} & 14:12:08.9 & +42:16:24.6 & $18.46\pm0.02$ & $-80.3\pm3.1$ & $-57.5\pm3.1$ & DA & \multirow{2}{*}{13.6} & \multirow{2}{*}{$<$0.1\%} & 1,2 \\
 & 14:12:07.7 & +42:16:27.1 & $15.83\pm0.01$ & $-81.7\pm2.7$ & $-61.4\pm2.7$ & DA & & & 4,6 \B \\
\hline
\T
\multirow{2}{*}{J1703$+$3304} & 17:03:55.9 & +33:04:38.3 & $18.81\pm0.02$ & $-1.8\pm3.4$ & $-51.2\pm3.4$ & DA & \multirow{2}{*}{11.2} & \multirow{2}{*}{$<$0.1\%} & 1,2 \\
 & 17:03:56.9 & +33:04:35.8 & $18.16\pm0.01$ & $+0.3\pm3.1$ & $-50.5\pm3.1$ & & & & \B \\   
\hline
\T
\multirow{2}{*}{J2115$-$0741\tablenotemark{e}} & 21:15:07.4 & $-$07:41:51.5 & $17.47\pm0.02$ & $-25.2\pm2.9$ & $-117.2\pm2.9$ & DA & \multirow{2}{*}{17.0} & \multirow{2}{*}{$<$0.1\%} & 1,2 \\
 & 21:15:07.4 & $-$07:41:34.5 & $16.81\pm0.01$ & $-30.0\pm2.8$ & $-117.9\pm2.8$ & & & & \B \\  
\hline
\T
\multirow{2}{*}{J2326$-$0023\tablenotemark{e}} & 23:26:58.8 & $-$00:23:39.9 & $19.33\pm0.05$ & $+51.6\pm3.3$ & $-30.7\pm3.3$ & DA & \multirow{2}{*}{9.5} & \multirow{2}{*}{$<$0.1\%} & 1,2 \\
 & 23:26:59.3 & $-$00:23:48.1 & $17.49\pm0.02$ & $+53.0\pm2.8$ & $-28.8\pm2.8$ & DA & & & 1,2 
\enddata 
\tablerefs{1--\citet{eisenstein06b}, 2--Kleinman et al., 3--\citet{wegner87}, 4--\citet{mccook99}, 5--\citet{oswalt94}, 6--\citet{green86}} 
\tablenotetext{a}{The secondary in this pair has an unidentified SDSS spectra.} 
\tablenotetext{b}{J1054$+$5307 is a wide WDWD previously identified by \citet{mccook99}.} 
\tablenotetext{c}{The secondary in J1113$+$3238 is included in the hypervelocity star survey of \citet{lepine05}.} 
\tablenotetext{d}{The secondary in J1412$+$4216 is included in the hypervelocity star survey of \citet{brown07}.} 
\tablenotetext{e}{J2115$-$0741 and J2326$-$0023 were proposed as wide WDWDs by \citet{greaves05}.} 
\end{deluxetable*} 
 
\section{Results and Discussion}\label{res} 
\subsection{New WDWDs} 
Based on the tests described above, we set $\theta < 2\amin$ and $P_m < 1\%$ as our criteria for identifying true pairs. 
 
Of the candidate pairs with SDSS spectra for both WDs, J1011$+$2450 and J1555$+$0239 have $\theta > 2\amin$ and $P_m > 1\%$, and are therefore very likely to be random alignments. The distance and RV data in Table~\ref{tab:dist_rv} are consistent with this, as the disagreements in one or both measurements are significant. J0915$+$0947, for which $\theta = 4.5\amin$, does have $P_m < 1\%$, but the RV measurements for the two WDs are highly discrepant ($-58\pm8$ and $+18\pm7$ km/s), implying that this is a random match. 
 
Conversely, five pairs listed in Table~\ref{tab:dist_rv} have $\theta < 2\amin$ and $P_m < 1\%$: J0332$-$0049, J1113$+$3238, J1257$+$1925, J1309$+$5503, and J2326$-$0023. We check the distances to the two components and the RV measurements for these for consistency: of the five, J1309$+$5503 is the only one for which the disagreements between the measurements for each DA ($-50\pm23$ and $+45\pm15$ km/s) are significant enough to eliminate it from our list of candidate pairs. We identify the other four pairs as high-confidence candidate WDWD systems and present them along with the eight systems that lack SDSS spectra for the secondaries but also meet our criteria in Table~\ref{tab:results}. 
  
\begin{figure} 
\centerline{\includegraphics[width=0.76\columnwidth,angle=90]{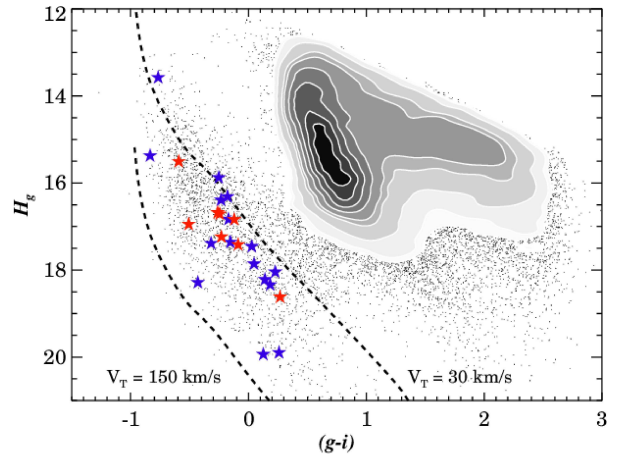}} 
\caption{Reduced proper motion as a function of $(g-i)$ for the SDSS stars presented in Figure~\ref{cmd} (points and contours) and for members of our high-confidence candidate WDWDs. Spectroscopically confirmed WDs are shown as blue stars, while the red stars lack spectra. The V$_{\rm T}=30$ km/s line marks the expected location of disk WDs and the V$_{\rm T}=150 $ km/s line that of halo WDs \citep{kilic06}. All of our candidates lie in the expected region for WDs, far from the main-sequence locus.}\label{rpm} 
\end{figure} 
 
In addition to the four pairs with SDSS spectra for both DAs, a SIMBAD\footnote{\tt http://simbad.u-strasbg.fr/simbad/} search finds that two systems in Table~\ref{tab:results} have secondaries classified as DAs by \citet{mccook99}. Spectroscopic follow-up is clearly needed to confirm the nature of the secondary in the remaining six systems. For objects $17 \leq g \leq 19$ mag, \citet{girven11} estimate that their photometric selection leads to a quasar contamination rate of $\lessim35\%$, primarily at the faint end. At the bright end, \citet{girven11} find that early-type main-sequence stars and subdwarfs are the main contaminants, and estimate their contamination rate to be $\lessim20\%$. 

However, the contamination rate drops sharply when a proper motion constraint is applied: if one imposes $\mu\ \gessim\ 20$ mas/yr, the quasar contamination rate becomes negligible and $\sim$90\% of the hot stars are eliminated. Considering that our high-confidence pairs all have $\mu > 30$ mas/yr, we expect that no more than one of our high-confidence pairs contains a non-WD. 

As a further test, we calculate the reduced proper motion ($H_g$) for each of our pairs. $H_g$, which combines photometric and kinematic information, is an effective tool for separating WDs from other objects \citep[][]{kilic06}. Figure~\ref{rpm} shows the reduced proper motions for the $>$99,000 SDSS stars presented in Figure~\ref{cmd}, as well as lines of constant transverse velocity representing the disk and halo WD populations (V$_{\rm T}=30$ and 150 km/s, respectively). All of the stars in our high-confidence pairs, whether spectroscopically confirmed as WDs or lacking spectra, lie far from the main-sequence locus and are consistent with being WDs.  
 
We note that two of these pairs, J2115$-$0741 and J2326$-$0023, were identified as wide WDWDs candidates by \citet{greaves05} in a search for CPMBs in SDSS Data Release 1. 
 
\begin{figure} 
\centerline{\includegraphics[width=0.7\columnwidth,angle=90]{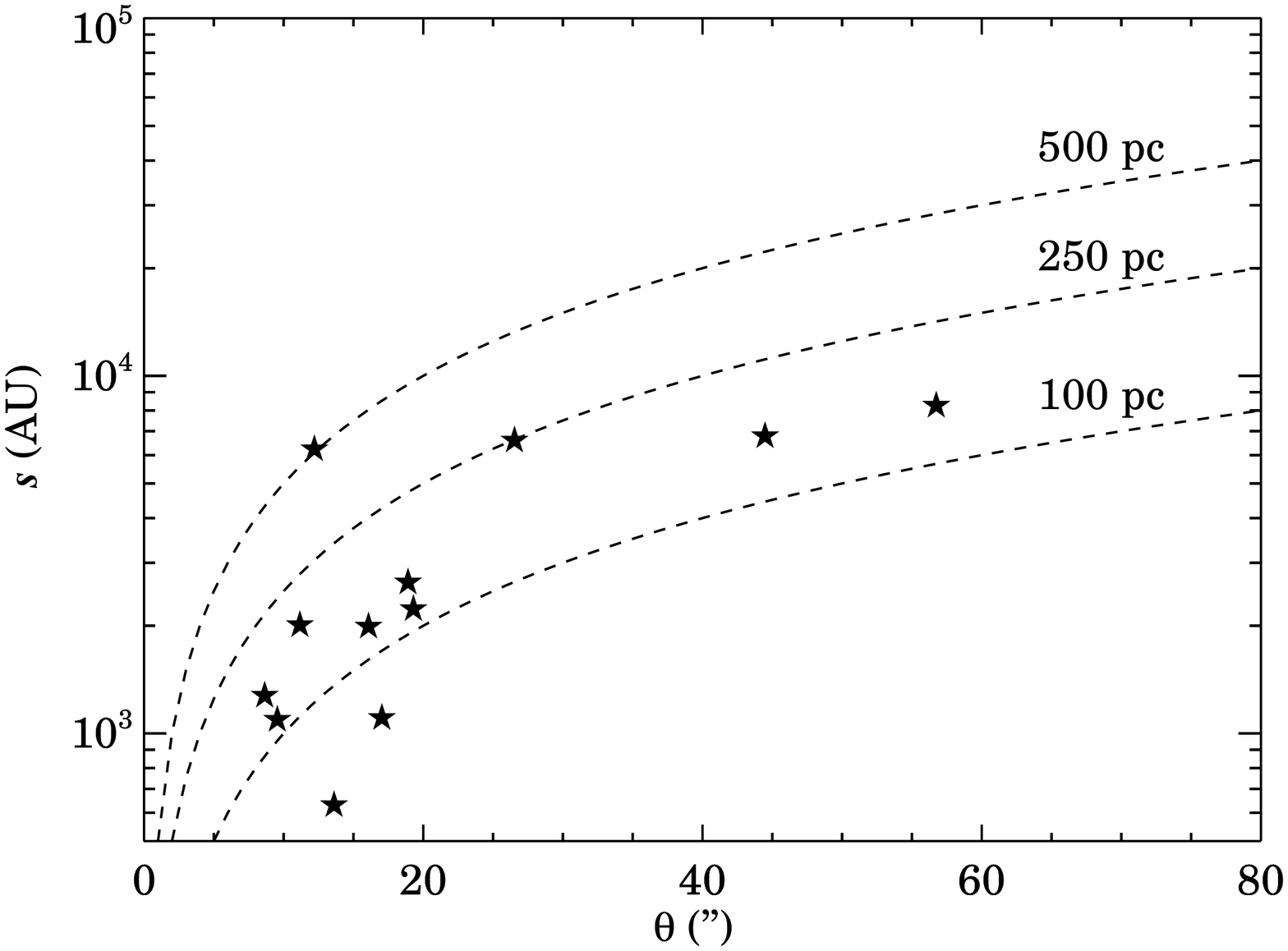}} 
\caption{Projected separation of our high-confidence WDWDs as a function of $\theta$. We use the spectroscopically determined distances to the primary WDs to estimate $s$. Lines of constant distance are plotted as dashed lines.}\label{separ} 
\end{figure}

In Figure \ref{separ}, we plot $s$, the projected orbital separation found using the spectroscopically derived distance to the primary WDs, as a function of $\theta$ for our high-confidence WDWDs. We find no pair with $s > 10^4$ AU, while our population synthesis predictions are that a significant population exists at such separations; we also appear only to be identifying very nearby WDWDs. However, our method cannot yield a complete sample: \citet{girven11} estimate the completeness of WD spectroscopic coverage in SDSS at only $\sim$44\% for $g \leq 19$ mag.

Our tightest pair has $\theta \sim 8\asec$, while we expect binaries to exist at smaller $\theta$. This is due to the difficulty in matching an object in the SDSS catalog to its USNO-B counterpart when the matching radius is of order the separation with another object. 

\begin{figure} 
\centerline{\includegraphics[width=0.7\columnwidth,angle=90]{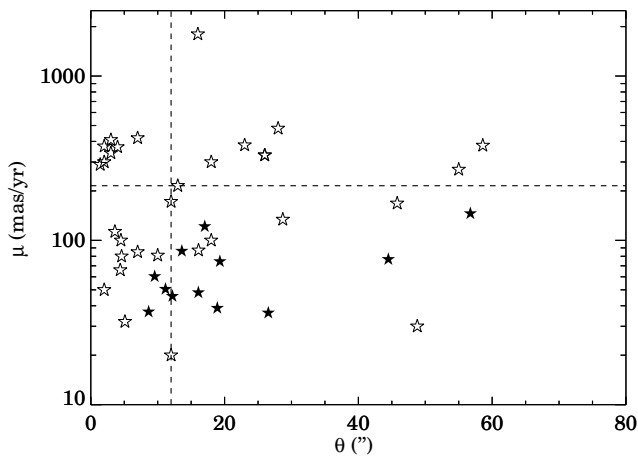}} 
\caption{WDWD proper motions as a function of $\theta$. Previously known systems are plotted as open stars, while our high-confidence pairs are filled stars. The dashed lines indicate the median $\mu$ and $\theta$ values for the previously known WDWDs; our high-confidence systems are a significant addition to the sample of pairs with small $\mu$ and large $\theta$.}\label{sepvspm} 
\end{figure} 
 
\tabletypesize{\tiny} 
\begin{deluxetable*}{llcccccccccl} 
\tablecaption{Wide WDWDs -- Derived Quantities \label{tab:known_sys}} 
\tablehead{ 
\colhead{} & 
\colhead{} & 
\colhead{} & 
\multicolumn{4}{c}{Primary} & 
\multicolumn{4}{c}{Secondary} & 
\colhead{} \\ 
\colhead{Name} & 
\colhead{$\theta$ ($\asec$)} & 
\colhead{$\mu$ (mas/yr)\tablenotemark{a}} & 
\multicolumn{1}{c}{$T_{\rm eff}$ (K)} & 
\colhead{log $g$} & 
\colhead{$M$ (\Msun)} & 
\colhead{log $\tau$} & 
\multicolumn{1}{c}{$T_{\rm eff}$ (K)} & 
\colhead{log $g$} & 
\colhead{$M$ (\Msun)} & 
\multicolumn{1}{c}{log $\tau$} & 
\colhead{Ref.}  
} 
 
\startdata 
\cutinhead{High-Confidence Wide WDWDs Identified in This Work\tablenotemark{b}} 
J0000$-$1051 & 16.1 & 49.7$\pm$8.7 & 8598$\pm$72 & 8.32$\pm$0.13 & 0.80$\pm$0.09 & 9.23$\pm$0.11 & & & & &  \\ 
J0029+0015 & 8.6 & 36.8$\pm$6.7 & 9947$\pm$67 & 8.22$\pm$0.08 & 0.74$\pm$0.06 & 8.97$\pm$0.08 & & & & & \\ 
J0332$-$0049\tablenotemark{c} & 18.9 & 36.5$\pm$9.9 & 11012$\pm$50 & 8.27$\pm$0.04 & 0.77$\pm$0.03 & 8.85$\pm$0.02 & 34288$\pm$42 & 7.83$\pm$0.02 & 0.58$\pm$0.01 & 6.60$\pm$0.01 & \\ 
J1002+3606 & 26.5 & 37.2$\pm$6.1 & 11326$\pm$181 & 8.05$\pm$0.13 & 0.63$\pm$0.08 & 8.68$\pm$0.09 & & & & & \\ 
J1054+5307\tablenotemark{c} & 44.5 & 77.6$\pm$5.0 & 10985$\pm$60 & 8.08$\pm$0.05 & 0.65$\pm$0.03 & 8.73$\pm$0.04 & 11120 & 8.01 & & & 1,2,3 \\ 
J1113+3238 & 56.7 & 147.3$\pm$5.8 & 6853$\pm$69 & 7.56$\pm$0.20 & 0.38$\pm$0.08 & 8.99$\pm$0.08 & 7580$\pm$88 & 8.40$\pm$0.15 & 0.85$\pm$0.10 & 9.46$\pm$0.09 & \\ 
J1203+4948 & 19.3 & 73.6$\pm$5.3 & 7064$\pm$46 & 8.05$\pm$0.10 & 0.62$\pm$0.06 & 9.21$\pm$0.08 & & & & & \\ 
J1257+1925 & 12.2 & 46.9$\pm$8.3 & 11829$\pm$218 & 7.72$\pm$0.14 & 0.46$\pm$0.06 & 8.46$\pm$0.05 & 47818$\pm$590 & 7.85$\pm$0.05 & 0.62$\pm$0.02 & 6.33$\pm$0.02 & \\ 
J1412+4216 & 13.6 & 84.4$\pm$5.1 & 6706$\pm$70 & 8.76$\pm$0.15 & 1.07$\pm$0.09 & 9.64$\pm$0.01 & & & & & \\ 
J1703+3304 & 11.2 & 50.9$\pm$6.0 & 9587$\pm$75 & 8.11$\pm$0.11 & 0.66$\pm$0.07 & 8.92$\pm$0.09 & & & & & \\ 
J2115$-$0741 & 17.0 & 120.7$\pm$5.6 & 7913$\pm$32 & 8.16$\pm$0.05 & 0.69$\pm$0.03 & 9.19$\pm$0.04 & & & & & \\ 
J2326$-$0023 & 9.5 & 60.2$\pm$6.1 & 7503$\pm$67 & 8.42$\pm$0.12 & 0.87$\pm$0.07 & 9.48$\pm$0.05 & 10513$\pm$46 & 8.24$\pm$0.05 & 0.75$\pm$0.04 & 8.90$\pm$0.04 & \\ 
\cutinhead{Previously Identified Wide WDWDs} 
LP406$-$62/63 & 28 & 480 & 5320 & 8.0 & 0.58 & 9.63 & 4910 & 8.0 & 0.58 & 9.78 & 1,2,4 \\ 
LP707$-$8/9 & 12 & 172 &  &  &  &  &  &  &  &  & 1 \\ 
LP647$-$33/34 & 2 & 374 &  &  &  &  &  &  &  &  & 1 \\ 
LP197$-$5/6 & 7 & 420 &  &  &  & 9.48 &  &  &  & 9.94 & 1,5 \\ 
RE J0317$-$853\tablenotemark{d} & 7 & 85 & 30000$-$50000 &  & >1.1 & $\sim$8.4 & 16000 & 8.19 & 0.76$-$0.84 & $\sim$8.4 & 6,7 \\ 
LP472$-$70/69\tablenotemark{e} & 3 &  &  &  &  &  &  &  &  &  & 1 \\ 
HS 0507+0434\tablenotemark{c} & 18 & 100 & 20220 & 7.99 & 0.62 & 7.90 & $\sim$12000 & 8.1 & 0.69 & 8.71 & 8 \\ 
WD 0727+482A/B\tablenotemark{e} &  & 1340 & 5020 & 7.92 & 0.53 & 9.66 & 5000 & 8.12 & 0.66 & 9.85 &  9 \\ 
LP543$-$33/32 & 16 & 1800 & 4170 & 7.65 & 0.39 & 9.67 & 4870 & 8.05 & 0.6 & 9.84 & 1,2,4,9 \\ 
LP035$-$288/287 & 3 & 340 &  &  &  &  &  &  &  &  & 1 \\ 
PG 0901+140 & 3.6 & 113 & 9500 & 8.29 &  & 9.10 & 8250 &  &  &  &  10,11 \\ 
PG 0922+162 & 4.4 & 66 & 22740 & 8.27 & 0.79 & 7.95 & 22130 & 8.78 & 1.1 & 8.41 & 10,12 \\ 
J0926+1321\tablenotemark{d} & 4.6 & 80 & 9500$\pm$500 &  & 0.62$\pm$0.10 & 8.86 & 10482$\pm$47 & 8.54$\pm$0.03 & 0.79$\pm$6 & 8.92 & 13 \\ 
LP462$-$56A/B & 4 & 370 & 10240 & 8.0 & 0.58 & 8.90 & 8340 & 7.5 & 0.35 & 8.77 & 1,2 \\ 
LP370$-$50/51 & 13 & 215 &  &  &  &  &  &  &  &  & 1 \\ 
LP549$-$33/32 & 26 & 330 &  &  &  &  &  &  &  &  & 1,3\\ 
PG 1017+125 & 48.8 & 30 &  &  &  &  &  &  &  &  & 10 \\ 
ESO439$-$162/163 & 23 & 380 & 5810 & 8.0 & 0.57 & 9.52 & 4780 & 8.0 & 0.57 & 9.82 & 1,2,14 \\ 
GD 322\tablenotemark{d} & 16.1 & 87 & 14790 & 7.87 & 0.54 & 8.26 & 6300 & 7.93 & 0.54 & 9.27 & 10,15 \\ 
LP322$-$500A/B & 12 & 20 &  &  &  &  &  &  &  &  & 1 \\ 
LP096$-$66/65 & 18 & 300 &  &  &  &  &  &  &  &  & 1,16 \\ 
L151$-$81A/B & 2 & 50 & 14050 & 7.96 & 0.57 & 8.38 & 12000 &  &  &  & 1,17,18,19 \\ 
J1507+5210\tablenotemark{d} & 5.1 & 32 & 17622$\pm$95 & 8.13$\pm$0.02 & 0.70$\pm$0.04 & 8.17 & 18000$\pm$1000 &  & 0.99$\pm$5 & 8.51 & 13 \\ 
Gr576/577\tablenotemark{f} & 4.5 & 100 & 12500 & 8.34 & 0.8 &  & 9500/8500 &  & 0.39/0.56 &   & 1,20,21,22 \\ 
LP567$-$39/38 & 2 & 300 &  &  &  &  &  &  &  &   & 1 \\ 
G206$-$17/18 & 55 & 270 & 7380 & 7.65 &  & 8.93 & 6480 & 7.75 &  & 9.12  & 1,9 \\ 
G021$-$15\tablenotemark{f} & 58.6 & 378 & 10000/15000 & 8.0/7.4 & 0.6/0.35 & 8.08 & 4750 & 8.0 & 0.57 &   & 9,10 \\ 
GD 392 & 45.8 & 168 & 12220 & 9.09 & 1.23 & 9.22 & $\sim$3600 &  &  &   & 10,19,23 \\ 
G261$-$43 & 1.4 & 289 & 16000 &  &  &  & 5000 &  &  &  & 10,24 \\ 
HS 2240+1234 & 10 & 81 & 14700 & 8.1 &  &  & 13200 & 7.9 &  &  & 8 \\ 
LP701$-$69/70 & 26 & 330 &  &  &  & 9.34 &  &  &  & 9.88 & 1,5 \\ 
GD 559 & 28.7 & 134 &  &  &  &  &  &  &  &  & 10,25\\ 
LP077$-$57/56 & 3 & 409 &  &  &  & 9.62 &  &  &  & 9.92 & 1,5  
\enddata 
 
\tablerefs{1--\citet{sion91}, 2--\citet{bergeron97}, 3--\citet{eisenstein06b}, 4--\citet{kilic09b}, 5--\citet{hintzen89}, 6--\citet{barstow95}, 7--\citet{kuelebi10}, 8--\citet{jordan98}, 9--\citet{bergeron01}, 10--\citet{farihi04b,farihi05}, 11--\citet{liebert05}, 12--\citet{finley97}, 13--\citet{dobbie11}, 14--\citet{ruiz95}, 15--\citet{girven10}, 16--\citet{kleinman04}, 17--\citet{oswalt88}, 18--\citet{wood92}, 19--\citet{bergeron11b}, 20--\citet{sanduleak82}, 21--\citet{greenstein83}, 22--\citet{maxted00}, 23--\citet{farihi04a}, 24--\citet{zuckerman97},  25--\citet{mccook99}} 
\tablenotetext{a}{As in the text, $\mu$ refers to the total proper motion of the system.} 
\tablenotetext{b}{The quoted values and uncertainties for $T_{\rm eff}$ and log $g$ are from the Kleinman et al.\ catalog. Uncertainties on the WD masses and cooling ages are formal and do not include significant systematic uncertainties.} 
\tablenotetext{c}{J0332$-$0049, J1054$+$5307, and HS 0507$+$0434 include a ZZ-Ceti-type variable.} 
\tablenotetext{d}{The primaries in RE J0317$-$853 and J0926$+$1321 and secondaries in GD 322 and J1507$+$5210 have been identified as magnetic WDs.} 
\tablenotetext{e}{LP472-70/69 lacks a published $\mu$, and WD 0727+482A/B lacks a published $\theta$.} 
\tablenotetext{f}{Gr576/577 and G021$-$15 have both been identified as triple degenerate systems.} 
\end{deluxetable*} 
 
\subsection{Comparison with Previously Known WDWDs} 
Table \ref{tab:known_sys} is a compilation of the properties of all of the wide WDWDs reported to date. We include our $12$ high-confidence WDWDs and $33$ systems from the literature and present $T_{\rm eff}$, log $g$, mass, and cooling age ($\tau$), when these measurements exist. The available data and their quality vary greatly from system to system, but simple comparisons can be made between our sample and the previously known WDWDs. In Figure~\ref{sepvspm}, we show $\mu$ as a function of $\theta$ for the previously known systems and for our pairs. Our high-confidence WDWDs are a significant addition to the sample of pairs with small $\mu$ ($<$200 mas/yr) and large $\theta$ ($>$10\asec). 

Interestingly, four of the systems listed in Table~\ref{tab:known_sys} include WDs with masses $\lessim 0.5\ \Msun$. The Galaxy is not thought to have had time to produce such low-mass WDs, as the youngest WDs in the oldest Galactic globular clusters have $M \sim 0.5\ \Msun$ \citep{hansen07}. Instead, WDs with $M < 0.5\ \Msun$ are likely to form in close binaries whose evolution included a phase of mass transfer. These four systems are therefore excellent candidate triple systems, with the low-mass WD likely to have a close-by companion. (Two additional systems in the literature are candidate triple systems.)

The previously known pairs include four WDs that are in the Kleinman et al.\ WD catalog. Of these, only LP 128$-$254/255 (J1054$+$5307) is recovered by our search; our primary in this case is LP 128$-$255. PG 0901$+$140 and J1507$+$5210 were excluded as candidate binaries because three of the four WDs in these pairs lack proper motion information in SDSS. LP 549$-$33/32 is not recovered because the secondary (in our case, LP 549$-$32) has colors outside the regions defined by \citet{girven11} for DA WDs; indeed, \cite{mccook99} classify it as a DC WD. The non-detection of three of these four systems is therefore not surprising, but confirms that our reliance on the Kleinman et al.\ DA catalog leads us to miss a number of wide WDWDs in the SDSS footprint. 
 
\subsubsection{Binary Separation and Stability} 
\begin{figure}
\centerline{\includegraphics[width=0.76\columnwidth,angle=90]{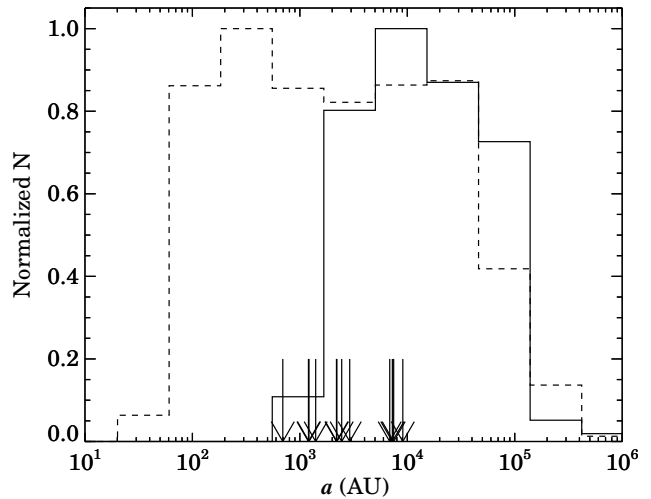}}
\caption{Normalized orbital separation distributions for the WDWDs generated by our population synthesis (dashed line) and for the wide main-sequence binaries identified by \citet[][solid line]{dhital10}. For the latter, we use $a = 1.1 \times s$ to convert the observed (projected) orbital separations into true orbital separations \citep{dupuy11}. The synthesized systems are those that remain after perturbing effects are taken into account (dashed line in Figure~\ref{Phist}). The projected separations for the 12 high-confidence WDWDs identified here have also been converted; their orbital separations are indicated by the arrows. }\label{sep_comp}
\end{figure}

As mentioned in Section \ref{motiv}, wide WDWD orbits are thought to be $\sim$5$\times$ larger than those of their progenitor systems; mass lost as the stars evolve expands the orbits from a more compact state \citep{greenstein86}. We compare the projected orbital separations for the WDWDs produced by our population synthesis to those for the wide main-sequence pairs identified by \citet{dhital10} in Figure~\ref{sep_comp}. (We also show the individual values for our 12 WDWDs; this sample is too small and incomplete for a comparison to the separation distribution to be meaningful.)

The apparent deficit of progenitor pairs for the synthesized WDWDs with $10^2\ \lessim\ a\ \lessim\ 10^3$ AU is unsurprising, as those progenitors presumably are tighter than the tightest pairs to which \citet{dhital10} are sensitive ($\theta \sim 8\asec$). More interesting is the apparent lack of a significant population of WDWDs wider than the widest main-sequence pairs identified by \citet{dhital10}. This may be because as the widest main-sequence pairs evolve to larger separations, they become more likely to be disrupted by interactions with other bodies in the Galaxy. 

One test of this hypothesis is to compare the ages of our WDWDs to those of the \citet{dhital10} binaries. We use the cooling ages of our primaries as a rough estimate of the cooling ages of the systems (we do not know whether they evolved first). This cooling age provides an upper limit on the WDWDs' lifetimes as truly wide binaries, since it is only after both stars have evolved into WDs that the pairs reach maximum separations. The characteristic lifetimes of the WDWDs in Table~\ref{tab:known_sys} derived in this manner are all $>>$1 Gyr. By contrast, \citet{dhital10} find that many of their pairs have characteristic lifetimes $\lessim$1 Gyr. A larger sample of WDWDs will allow for an improved test of this hypothesis.
 
\subsubsection{Initial-Final Mass Relation}  
While the IFMR is often studied using WDs in open clusters \citep[e.g.,][]{kalirai2008} or in binaries in which the WD has a main-sequence companion \citep[e.g.,][]{catalan2008a,zhao12}, wide WDWDs can also be used to constrain this relation. \citet{finley97} estimated the initial mass of the massive secondary in the binary PG0922$+$162 by comparing it to similarly massive WDs in open clusters with well-known ages. Subtracting the cooling age from the host cluster ages gave the pre-WD lifetime of these reference WDs and therefore of the secondary in PG 0922$+$162. This was added to the secondary's cooling age to obtain the total age of the system. \citet{finley97} were then able to use stellar models to estimate the primary's initial mass. The resulting data point in the initial-final mass plane has relatively small error bars and can be used to discriminate between different theoretical IFMRs.  
 
However, most of the 45 systems in Table \ref{tab:known_sys} lack spectroscopic information, and many of the WDs with spectra have large error bars on $T_{\rm eff}$ or log $g$ because these were derived from low signal-to-noise spectra. Such errors can propagate into large uncertainties in the masses and cooling ages. While in principle these wide WDWDs can be used to constrain the IFMR with a method similar to that of \citet{finley97}, we defer a full study of this question to a later paper.   Spectroscopic follow-up of an expanded set of WDWDs, combined with ever improving WD atmospheric models, may eventually produce a large, homogeneous dataset with which to constrain the IFMR. 
 
For now, as a simple consistency check, we compare the cooling ages and masses of the WDs in the eight systems for which both stars have spectroscopically derived $T_{\rm eff}$ and differing masses (and that are not a potential triple system). We expect that in these binaries the more massive WD has a larger cooling age. Our simple test holds for all eight of these systems.  
  
\section{Conclusions} 
To expand the sample of known WDWDs, we adapted the method of \citet{dhital10} to search the SDSS DR7 catalog for WD common proper motion companions to spectroscopically confirmed DA WDs out to $\theta = 10\amin$. We found 41 candidate wide WDWDs. These are pairs in which the secondary's colors fall within 0.5$\sigma$ of the region in $(g-r)$ versus $(u-g)$ color space occupied by DA WDs \citep{girven11}. We then used two complementary techniques to measure the contamination of our sample by randomly aligned false matches. We first estimated the overall contamination of our sample by false binaries by shifting the positions of our primaries several times by 1$^\circ$ and applying our $\mu$ and color criteria to identify false companions to the shifted primaries. We also used the Monte Carlo approach developed by \citet{dhital10} and searched $10^5$ iterations of the line-of-sight to each of our primaries for stars whose properties randomly match those of these WDs. 
 
These two tests suggest that, absent other information, the highest probability real pairs in our sample are those with $\theta <2\amin$ and primaries with random matches in fewer than 1\% of our rendered lines-of-sight. 13 of our candidates meet these criteria; one of these pairs has inconsistent spectroscopically derived distance and RV measurements for the two WDs, so that our final sample includes 12 high-confidence wide WDWDs. (Three of these were previously reported as candidate WDWDs.) 
 
Four of these pairs have SDSS spectra for both DAs, while two others have secondaries classified as DAs by \citet{mccook99}. Spectroscopic follow-up is clearly needed to confirm the nature of the secondaries in the remaining six systems. However, a color selection combined with a minimum proper motion requirement returns a very clean sample of DAs \citep{girven11}; furthermore, based on their reduced proper motions, all of our candidates are consistent with being WDs. We therefore expect no more than one of our high-confidence pairs to contain a non-WD. These systems are a significant addition to the known population with small $\mu$ ($<$200 mas/yr) and large $\theta$ ($>$10\asec).  
 
\citet{girven11} estimate the completeness of WD spectroscopic coverage in SDSS at $\sim$44\% for $g \leq 19$ mag, and it decreases for fainter magnitudes. This is particularly unfortunate as the SDSS proper motion catalog goes to $g \sim 20$.  Extending the method developed to identify the 12 pairs presented here to a photometrically selected set of primary WDs could significantly increase the number of known WDWDs, which is currently $<$50. An expanded set of WDWDs and spectroscopic follow-up may eventually produce a large enough dataset for these pairs to realize their full potential as testbeds for theories of stellar evolution.   
 
\acknowledgments 
We are grateful to the anonymous referee whose comments improved this work. We thank Sebastian L\'epine and Marten van Kerkwijk for useful discussions about identifying false pairs. We thank Kepler Oliveira for obtaining $T_{\rm eff}$ and log $g$ for a WD primary whose data in our catalog were unusable. We also thank Mukremin Kilic for his help with Figure~\ref{rpm}. 
 
This research has made use of the SIMBAD database, operated at CDS, Strasbourg, France. 
 
Funding for the SDSS and SDSS-II has been provided by the Alfred P. Sloan Foundation, the Participating Institutions, the National Science Foundation, the U.S. Department of Energy, the National Aeronautics and Space Administration, the Japanese Monbukagakusho, the Max Planck Society, and the Higher Education Funding Council for England. The SDSS Web Site is {\tt http://www.sdss.org/}. 
 
The SDSS is managed by the Astrophysical Research Consortium for the Participating Institutions. The Participating Institutions are the American Museum of Natural History, Astrophysical Institute Potsdam, University of Basel, University of Cambridge, Case Western Reserve University, University of Chicago, Drexel University, Fermilab, the Institute for Advanced Study, the Japan Participation Group, Johns Hopkins University, the Joint Institute for Nuclear Astrophysics, the Kavli Institute for Particle Astrophysics and Cosmology, the Korean Scientist Group, the Chinese Academy of Sciences (LAMOST), Los Alamos National Laboratory, the Max-Planck-Institute for Astronomy (MPIA), the Max-Planck-Institute for Astrophysics (MPA), New Mexico State University, Ohio State University, University of Pittsburgh, University of Portsmouth, Princeton University, the United States Naval Observatory, and the University of Washington.

\end{document}